\documentclass[aps,preprintnumbers,showpacs, nofootinbibt,referee,keywords,twocolumn]{revtex4}
\usepackage{graphicx}
\usepackage{amsmath}
\usepackage{bm}
\usepackage{color}

\def\be{\begin{equation}}
\def\ee{\end{equation}}
\def\bea{\begin{eqnarray}}
\def\eea{\end{eqnarray}}

\begin{document}

\title{Structure of neutron, quark and exotic stars in Eddington-inspired
Born-Infeld gravity}
\author{Tiberiu Harko$^1$}
\email{t.harko@ucl.ac.uk}
\author{Francisco S.~N.~Lobo$^{2}$}
\email{flobo@cii.fc.ul.pt}
\author{M. K. Mak$^{3}$}
\email{mkmak@vtc.edu.hk}
\author{Sergey V. Sushkov$^{4}$}
\email{sergey_sushkov@mail.ru}
\affiliation{$^1$Department of Mathematics, University College London, Gower Street,
London WC1E 6BT, United Kingdom}
\affiliation{$^2$Centro de Astronomia e Astrof\'{\i}sica da Universidade de Lisboa, Campo
Grande, Ed. C8 1749-016 Lisboa, Portugal}
\affiliation{$^{3}$Department of Computing and Information Management, Hong Kong
Institute of Vocational Education, Chai Wan, Hong Kong, P. R. China,}
\affiliation{$^4$Institute of Physics, Kazan Federal University, Kremlevskaya Street 18,
Kazan 420008, Russia}
\date{\today}

\begin{abstract}
We consider the structure and physical properties of specific classes of
neutron, quark and ``exotic'' stars in Eddington-inspired Born-Infeld (EiBI)
gravity. The latter reduces to standard general relativity in vacuum, but
presents a different behavior of the gravitational field in the presence of
matter. The equilibrium equations for a spherically symmetric configuration
(mass continuity and Tolman-Oppenheimer-Volkoff) are derived, and their
solutions are obtained numerically for different equations of state of
neutron and quark matter. More specifically, stellar models, described by the stiff fluid,
radiation-like, polytropic and the bag model quark equations of state are
explicitly constructed in both general relativity and EiBI gravity, thus
allowing a comparison between the predictions of these two gravitational models.
As a general result it turns out that for all the considered equations of
state, EiBI gravity stars are more massive than their general relativistic
counterparts. Furthermore, an exact solution of the spherically symmetric
field equations in EiBI gravity, describing an ``exotic'' star, with
decreasing pressure but increasing energy density, is also obtained. As a
possible astrophysical application of the obtained results we suggest that
stellar mass black holes, with masses in the range of $3.8M_{\odot}$ and $%
6M_{\odot}$, respectively, could be in fact EiBI neutron or quark stars.
\end{abstract}

\maketitle



\section{Introduction}

Despite its remarkable successes, the standard $\Lambda $CDM ($\Lambda $
Cold Dark Matter) cosmological model faces severe theoretical,
interpretational and observational challenges. The most important of these is
the explanation of the late accelerated expansion of the Universe, inferred from
observations of the expansionary evolution of Type Ia supernovae \cite{Riess}.
Combined with the recent Cosmic Microwave Background observations of the
Planck satellite, \cite{Planck}, astronomical and astrophysical data provide
compelling evidence that our Universe is dominated by a mysterious and
exotic component, whose properties are difficult to be understood in the
framework of our present day knowledge. Indeed, the standard model of
cosmology has favored a missing energy-momentum component, in particular,
the dark energy models. This exotic component can be interpreted
theoretically either by assuming that it is a cosmological constant, which
would represent an intrinsic curvature of space-time, or a vacuum energy.
Alternatively, the dominant component of the Universe can be seen as a dark
energy, which would mimic a cosmological constant. One of main dark energy
scenarios is based on the so-called quintessence, where dark energy
corresponds to a dynamical scalar field $\phi$ \cite{quint}.

On the other hand, the possibility that general relativity breaks down at
cosmological scales cannot be ruled out {\it a priori}, and the late-time cosmic
acceleration may be due to infra-red modifications of general relativity.
Therefore, a second possibility in explaining the observational data is to
assume that at large scales the nature of the gravitational interaction is
modified, and a new theoretical model of gravity is necessary in order to
understand, and interpret, the observational data. Several, essentially
geometric, modifications of standard general relativity have been
considered, and investigated in detail as alternatives to dark energy. In
particular, the $f(R)$ type models \cite{rev}, where $R$ is the Ricci
scalar, $f\left( R,L_{m}\right) $ models with geometry-matter coupling \cite%
{coupl}, where $L_{m}$ is the matter Lagrangian, $f(R,T)$ models \cite{fRT},
where $T$ is the trace of the energy-momentum tensor, Weyl-Cartan-Weitzenb%
\"{o}ck gravity \cite{WCW}, hybrid metric-Palatini $f(X)$-gravity models %
\cite{fX}, or the recently proposed $f\left( R,T,R_{\mu \nu }T^{\mu \nu }\right) $ gravity \cite%
{fRTT}, where $R_{\mu \nu }$ is the Ricci tensor, and $T_{\mu \nu }$ is the
matter energy-momentum tensor, are some of the proposed geometric
modifications of general relativity that can explain the late de Sitter type
expansionary phase in the evolution of the Universe.

In the context of modified theories of gravity,
based on the classic work of Eddington \cite{Ed}, and on the non-linear
electrodynamics of Born and Infeld \cite{BI}, an interesting extension of
general relativity was introduced in \cite{Deser}, and further developed in %
\cite{Ban}. Essentially, in this model, denoted Eddington-inspired
Born-Infeld gravity (EiBI), the Eddington action is coupled to matter
without insisting on a purely affine action, or on a theory equivalent to
Einstein gravity. The metric is present in the model, and the gravitational
action has a Born-Infeld like structure. The model is based on a
Palatini-type formulation, with the metric tensor $g_{\mu \nu}$ and the
connection $\Gamma _{\alpha \beta }^{\mu }$ are varied independently. In this
model, the Newton-Poisson equation is modified in the presence of matter
sources, and the charged black holes are similar with those arising in
Born-Infeld electrodynamics coupled to gravity. The cosmological solutions
of the model for homogeneous and isotropic space-times show that there is a
minimum length (and maximum density) at early times, indicating the
possibility of an alternative theory of the Big Bang \cite{Ban}. For a
positive coupling parameter, the field equations have an important impact on
the collapse of dust, and do not lead to singularities \cite{P}. The theory
supports stable, compact pressureless stars made of perfect fluid, and the
existence of relativistic stars imposes a strong, near optimal constraint on
the coupling parameter. This constraint can be improved by observations of
the moment of inertia of double pulsars.

In \cite{T} it was shown that
the EiBI theory coupled to a perfect fluid reduces to general relativity
coupled to a nonlinearly modified perfect fluid, leading to an ambiguity
between the modified coupling and the modified equation of state.
The observational consequences of this degeneracy were discussed, and it was
argued that this extension of general relativity is viable from both an
experimental and theoretical point of view through the energy conditions,
consistency, and singularity-avoidance perspectives \cite{T}. However, in \cite{Sot}
it was shown that the EiBI theory, which is reminiscent of Palatini $f(R)$
gravity, shares the same pathologies, such as curvature
singularities at the surface of polytropic stars and unacceptable Newtonian
limit. The singularity avoidance in EiBI gravity was analyzed in \cite{Lop},
by considering the behavior of a homogeneous and isotropic universe filled
with phantom energy in addition to the dark and baryonic matter. Unlike the
Big Bang singularity that can be avoided in this kind of model, the Big Rip
singularity is unavoidable in the EiBI phantom model. The dark matter
density profile in EiBI gravity was also considered in \cite{dark}, and it was
found that in this model the dark matter density distribution is described
by the Lane-Emden equation with a polytropic index $n=1$, and is non-singular
at the galactic center. The tensor perturbations of a homogeneous and
isotropic space-time in the Eddington regime, where modifications to
Einstein gravity are strong were analyzed, and it was found that the
tensor mode is linearly unstable deep in the Eddington regime \cite%
{EscamillaRivera:2012vz}. Furthermore, it was also argued that EiBI cosmologies may present
viable alternatives to the inflationary paradigm as a solution to fundamental
problems of the standard cosmological model, and that under specific
assumptions the model is free from tensor singularities \cite{Avelino:2012ue}.
Other cosmological and astrophysical aspects of the EiBI model were
considered in \cite{all}.

In an astrophysical context, the hydrostatic equilibrium
structure of compact stars in the EiBI gravity was explored in \cite{Lin1},
and a framework to study the radial perturbations and stability of compact
stars in this theory was also developed. The standard results of stellar
stability still hold in the EiBI theory, with the frequency square of the
fundamental oscillation mode vanishing for the maximum-mass stellar
configuration. The dependence of the oscillation mode frequencies on the
coupling parameter $\kappa$ of the theory was also investigated. The
fundamental mode is insensitive to the value of the coupling constant, while
higher order modes depend more strongly on it. However, generic phase transitions
taking place in compact stars constructed in the framework of the EiBI
gravity can lead to anomalous behavior of these stars \cite{Lin2}. In the
case of first-order phase transitions, compact stars in EiBI gravity with a
positive coupling parameter $\kappa $ possess a constant pressure finite
region, which is not present in general relativistic stars. For the case of
a negative $\kappa $, an equilibrium stellar configuration cannot be
realized. Hence, in EiBI gravity there are stricter constraints on the
microphysics of the stellar matter. Besides, in the presence of spatial
discontinuities in the speed of sound due to phase transitions, the Ricci
scalar is spatially discontinuous, and contains delta-function
singularities, proportional to the jump in the speed of sound \cite{Lin2}.

It is the purpose of this paper to investigate the properties of
relativistic compact stars in the EiBI model. By assuming a spherically
symmetric perfect fluid matter, the gravitational field equations of the
EiBI model are solved numerically with several prescribed equations of
state. As specific examples of stellar models we consider stars described by
the causal stiff fluid equation of state, for which the speed of sound
equals the speed of light; the radiation-type equation of state, for which
the trace of the energy-momentum tensor is zero; the degenerate relativistic
neutron matter equation of state, representing a polytrope of index $n=3$;
and the quark matter equation of state. For all these models the global
astrophysical parameters of the stars (radius and mass) are obtained in both
standard general relativity and in the EiBI gravity model, thus allowing a
detailed comparison of the two approaches to stellar structure. As a general
result of our study it follows that EiBI gravity allows the existence of
more massive stars, as compared to general relativity. We also obtain an
exact stellar model solution, corresponding to an equation of state of the form $\rho
+3p=1/4\pi \kappa $, where $\rho $ and $p$ are the energy density and
isotropic pressure, respectively. This model corresponds to an ``exotic''
EiBI stellar-type object, with decreasing pressure, but increasing energy
density.

The present paper is organized as follows. The EiBI gravity theory is briefly
presented in Section \ref{sect2}. The system of gravitational field
equations describing the star interior (mass continuity and hydrostatic
equilibrium equations) is derived in Section~\ref{sect3}. Stellar models
described by the stiff fluid, radiation, polytropic and MIT bag model
equations of state are studied numerically, in both EiBI model and standard
general relativity, in Section~\ref{sect4}. An exact ``exotic'' stellar
model is obtained in Section~\ref{sect5}. Finally, we discuss and conclude
our results in Section~\ref{sect6}.

\section{Eddington-Inspired Born-Infeld gravity: Formalism}
\label{sect2}

In the present Section, we adopt for simplicity the natural system of units
with $G=c=1.$ The EiBI theory, which is based on the Eddington gravitational
action \cite{Ed} and Born-Infeld nonlinear electrodynamics \cite{BI}, is
obtained from the action $S$ given by \cite{Ban}
\begin{eqnarray}
S &=&\frac{1}{16\pi }\frac{2}{\kappa }\int d^{4}x\left( \sqrt{-\left| g_{\mu
\nu }+\kappa R_{\mu \nu }\right| }-\lambda \sqrt{-g}\right)  \notag
\label{1c} \\
&&+S_{M}\left[ g,\Psi _{M}\right] ,
\end{eqnarray}%
where $g=\mathrm{det}(g_{\mu \nu })$ and $R_{\mu \nu }$ is the symmetric
part of the Ricci tensor, which is constructed solely from the connection $%
\Gamma _{\beta \gamma }^{\alpha }$. The determinant of the tensor $g_{\mu
\nu }+\kappa R_{\mu \nu }$ is denoted by $\left| g_{\mu \nu }+\kappa R_{\mu
\nu }\right| $. In addition to this, $\lambda \neq 0$ is a dimensionless
constant and $\kappa $ is the Eddington parameter with inverse dimension to
that of the cosmological constant $\Lambda $.

The matter action $S_{M}$ depends only on the metric $g_{\mu \nu }$ and the
matter fields $\Psi _{M}$. In the limit $\kappa \rightarrow 0$, the action (%
\ref{1c}) recovers the Einstein-Hilbert action with $\lambda =\Lambda \kappa
+1$. In the present paper, we consider only asymptotic flat solutions, and
hence we take $\lambda =1$. Therefore the cosmological constant vanishes,
and the remaining parameter $\kappa $ plays the fundamental role for
describing the physical behavior of various cosmological and stellar
scenarios. Several constraints on the value and the sign of the parameter $%
\kappa $ have been obtained from solar observations, big bang
nucleosynthesis, and the existence of neutron stars in \cite{Ban,P,k3,k4}.
In particular, for cases with positive $\kappa $, effective gravitational
repulsion prevails, leading to the existence of pressureless stars (stars
made of non-interacting particles which provide interesting models for
self-gravitating dark matter \cite{dark}) and to an increase in the mass
limits of compact stars \cite{P,Lin1}.

Note that in the EiBI theory the metric $g_{\mu \nu }$ and the connection $%
\Gamma _{\beta \gamma }^{\alpha }$ are treated as independent fields.
Variation of the action (\ref{1c}) leads to the following results \cite{Ban,T,DN}:
\begin{eqnarray}
q_{\mu \nu }&=&g_{\mu \nu }+\kappa R_{\mu \nu },  \label{2c} \\
q^{\mu \nu }&=&\tau \left( g^{\mu \nu }-8\pi \kappa T^{\mu \nu }\right) ,
\label{3c} \\
\Gamma _{\beta \gamma }^{\alpha }&=&\frac{1}{2}q^{\alpha \sigma }\left(
\partial _{\gamma }q_{\sigma \beta }+\partial _{\beta }q_{\sigma \gamma
}-\partial _{\sigma }q_{\beta \gamma }\right) ,  \label{4c}
\end{eqnarray}%
where $q_{\mu \nu }$ is an auxiliary metric, $q=\mathrm{det}(q_{\mu \nu })$
and we have denoted $\tau $ as $\tau =\sqrt{g/q}$.

In the EiBI model, the energy-momentum tensor $T^{\mu \nu }$, defined as
\begin{equation}
T^{\mu \nu }=\frac{1}{\sqrt{-g}}\frac{\delta S_{M}}{\delta g_{\mu \nu }},
\end{equation}%
satisfies the standard conservation equations $\nabla _{\mu }T^{\mu \nu
}=0$, where, as in general relativity, the covariant derivative $\nabla
_{\mu }$ refers to the metric $g_{\mu \nu }$. If the energy-momentum tensor $%
T^{\mu \nu }$ vanishes in Eq.~(\ref{3c}), then the physical metric $g_{\mu
\nu }$ is equal to the apparent metric $q_{\mu \nu }$. Hence in vacuum the
EiBI theory is completely equivalent to standard general relativity.

Note that Eqs. (\ref{2c}) and (\ref{3c}) may be expressed in the following
forms \cite{T}
\begin{eqnarray}
q^{\mu \alpha }g_{\alpha \nu } &=&\delta ^{\mu }{}_{\nu }-\kappa R^{\mu
}{}_{\nu },  \label{field1} \\
q^{\mu \alpha }g_{\alpha \nu } &=&\tau \left( \delta ^{\mu }{}_{\nu }-8\pi
\kappa T^{\mu }{}_{\nu }\right) ,  \label{field2}
\end{eqnarray}%
where $R^{\mu }{}_{\nu }=q^{\mu \alpha }R_{\alpha \nu }$ and $T^{\mu
}{}_{\nu }=g^{\mu \alpha }T_{\alpha \nu }$. Now, combining Eqs. (\ref{field1}%
)-(\ref{field2}), yields the following relations
\begin{eqnarray}
R^{\mu }{}_{\nu } &=&8\pi \tau T^{\mu }{}_{\nu }+\frac{1-\tau }{\kappa }%
\delta ^{\mu }{}_{\nu },  \label{field1b} \\
R &=&8\pi \tau T+\frac{4(1-\tau )}{\kappa }.  \label{field2b}
\end{eqnarray}%
One may now write the modified Einstein equation as
\begin{equation}
G^{\mu }{}_{\nu }\equiv R^{\mu }{}_{\nu }-\frac{1}{2}R\delta ^{\mu }{}_{\nu
}=8\pi \tau T^{\mu }{}_{\nu }-\left( \frac{1-\tau }{\kappa }+4\pi \tau
T\right) \delta ^{\mu }{}_{\nu },  \label{modEFE}
\end{equation}%
where the Einstein tensor $G^{\mu }{}_{\nu }$ is defined in terms of the
auxiliary $q$-metric. The factor $\tau $ can be obtained from $T^{\mu
}{}_{\nu }$ by the relation
\begin{equation}
\tau =\left[ \mathrm{det}(\delta ^{\mu }{}_{\nu }-8\pi \kappa T^{\mu
}{}_{\nu })\right] ^{-\frac{1}{2}}.
\end{equation}

Throughout this work we consider that the energy-momentum tensor of the
compact object is given by the standard form
\begin{equation}
T^{\mu \nu }=\left( \rho +p\right) u^{\mu }u^{\nu }+pg^{\mu \nu },
\end{equation}%
where $\rho $, $p$ and $u^{\mu }$ are the energy density, the isotropic
pressure and the four velocity of the fluid, respectively, with the latter
satisfying the normalization condition $u^{\mu }u^{\nu }g_{\mu \nu }=-1$.
Thus, in terms of physical quantities $\tau $ can be expressed as
\begin{equation}
\tau =\left[ \left( 1+8\pi \kappa \rho \right) \left( 1-8\pi \kappa p\right)
^{3}\right] ^{-\frac{1}{2}}.  \label{19c}
\end{equation}

With the EiBI gravity theory briefly presented above, we now analyze the structure equations for static and spherically symmetric compact objects below.

\section{Structure equations for compact objects in Eddington-inspired
Born-Infeld gravity}

\label{sect3}

In the following, we shall incorporate the natural system of units $G$ and $%
c $ to the corresponding equations. Now, we will investigate the structure
of compact static and spherically symmetric objects. The line elements for
the physical metric $g_{\mu \nu }$ and for the auxiliary metric $q_{\mu \nu
} $ are given by \cite{Lin2}
\begin{eqnarray}
g_{\mu \nu }dx^{\mu }dx^{\nu } &=&-e^{\nu \left( r\right)
}c^{2}dt^{2}+e^{\lambda \left( r\right) }dr^{2}+f\left( r\right) d\Omega
^{2},  \label{8c} \\
q_{\mu \nu }dx^{\mu }dx^{\nu } &=&-e^{\beta \left( r\right)
}c^{2}dt^{2}+e^{\alpha \left( r\right) }dr^{2}+r^{2}d\Omega ^{2},  \label{9c}
\end{eqnarray}%
respectively, where $\nu(r)$, $\lambda(r)$, $\beta(r)$, $\alpha(r)$ and $%
f(r) $ are arbitrary metric functions of the radial coordinate $r$, and $%
d\Omega ^{2}=d\theta ^{2}+\sin ^{2}\theta d\phi ^{2}$.

Using Eq. (\ref{modEFE}), the system of gravitational field equations
describing the structure of a compact object is given by \cite{Lin1,Lin2}
\begin{eqnarray}
\frac{d}{dr}\left( re^{-\alpha }\right) =1-\frac{1}{2\kappa }\left( 2+\frac{a%
}{b^{3}}-\frac{3}{ab}\right) r^{2},  \label{g1} \\
e^{-\alpha }\left( 1+r\frac{d\beta }{dr}\right) =1+\frac{1}{2\kappa }\left(
\frac{1}{ab}+\frac{a}{b^{3}}-2\right) r^{2},  \label{g2}
\end{eqnarray}
and Eq. (\ref{3c}) yields the following relations
\begin{eqnarray}
e^{\beta }=\frac{e^{\nu }b^{3}}{a}, \qquad e^{\alpha }=e^{\lambda }ab,
\qquad f=\frac{r^{2}}{ab},  \label{x1}
\end{eqnarray}%
where we have defined the arbitrary functions $a\left( r\right) $ and $%
b\left( r\right) $ as
\begin{equation}
a=\sqrt{1+\frac{8\pi G}{c^{2}}\kappa \rho },
\end{equation}%
and
\begin{equation}
b=\sqrt{1-\frac{8\pi G}{c^{4}}\kappa p},
\end{equation}%
respectively.

The conservation of the energy-momentum tensor in the $g$-metric,
\begin{equation}
\frac{d\nu }{dr}=-\frac{2}{p+\rho c^{2}}\frac{dp}{dr}=\frac{4b}{a^{2}-b^{2}}%
\frac{db}{dr},  \label{psi}
\end{equation}%
provides the following conservation relation in the auxiliary $q$-metric
\begin{equation}
\frac{d\beta }{dr}=\frac{4b}{a^{2}-b^{2}}\frac{db}{dr}+\frac{3}{b}\frac{db}{%
dr}-\frac{1}{a}\frac{da}{dr}.  \label{L}
\end{equation}
The existence of a barotropic equation of state of the dense matter $%
p=p(\rho )$ imposes a similar equation of state in the $q$-metric, $a=a(b)$.
Therefore, by defining $c_{q}^{2}=da(b)/db$, the
energy-momentum conservation equation in the $q$-metric can be formulated as
\begin{equation}
\frac{d\beta }{dr}=\left( \frac{4b}{a^{2}-b^{2}}+\frac{3}{b}-\frac{1}{a}%
c_{q}^{2}\right) \frac{db}{dr}.  \label{betaprime}
\end{equation}

Note that Eq.~(\ref{g1}) can be immediately integrated to give
\begin{equation}  \label{alpha}
e^{-\alpha }=1-\frac{2Gm(r)}{c^2r},
\end{equation}
where the function $m(r)$ is obtained as
\begin{equation}  \label{masscont}
\frac{dm}{dr}=\frac{c^2}{4G\kappa }\left( 2+\frac{a}{b^{3}}-\frac{3}{ab}%
\right)r^2 .
\end{equation}

By substituting Eqs.~(\ref{betaprime})  and (\ref{alpha}) into Eq.~(\ref{g2})
we obtain the $q$-metric generalization of the standard
Tolman-Oppenheimer-Volkoff (TOV) equation of general relativity as
\begin{equation}
\frac{db}{dr}=\frac{ab\left( a^{2}-b^{2}\right) \left[ (1/2\kappa )\left(
1/ab+a/b^{3}-2\right) r^{3}+2Gm/c^{2}\right] }{r^{2}\left(
1-2Gm/c^{2}r\right) \left[ 4ab^{2}+3a\left( a^{2}-b^{2}\right) -b\left(
a^{2}-b^{2}\right) c_{q}^{2}\right] }.  \label{TOV}
\end{equation}

Once the equation of state of matter is known, the mass continuity, Eq.~(%
\ref{masscont}), and the generalized hydrostatic Eq.~(\ref{TOV}), describe
all the properties of compact objects in EiBI gravity. In order to obtain a
dimensionless form of the mass continuity and hydrostatic equilibrium
equations we introduce a set of dimensionless variables $\left( \eta
,m_{0},\kappa _{0},\theta ,p_{0}\right) $, defined as
\begin{eqnarray}  \label{var}
r &=&\frac{c}{\sqrt{2\pi G\rho _{c}}}\eta ,\qquad m=\frac{c^{3}}{\sqrt{2\pi
G^{3}\rho _{c}}}m_{0},  \notag  \label{dim} \\
\kappa &=&\frac{c^{2}}{8\pi G\rho _{c}}\kappa _{0},\qquad \rho =\rho
_{c}\theta ,\qquad p=\rho _{c}c^{2}p_{0},
\end{eqnarray}%
where $\rho _{c}$ is the central density of the star. These dimensionless quantities will be extremely useful for the numerical analysis carried out below.

Therefore in EiBI gravity the mass continuity and hydrostatic equilibrium
equations for compact objects take the dimensionless form
\begin{equation}
\frac{dm_{0}}{d\eta }=\frac{1}{\kappa _{0}}\left( 2+\frac{a^{2}-3b^{2}}{%
ab^{3}}\right) \eta ^{2},
\end{equation}%
and
\begin{equation}
\frac{db}{d\eta }=\frac{2\left[ \left( \frac{a^{2}+b^{2}}{ab^{3}}-2\right)
(\eta ^{3}/\kappa _{0})+m_{0}\right] }{\eta ^{2}\left( 1-2m_{0}/\eta \right) %
\left[ \frac{4b}{a^{2}-b^{2}}+\frac{3}{b}-\frac{d\ln a}{db}\right] },
\end{equation}%
respectively. The functions $a$ and $b$ are obtained as
\begin{equation}
a=\sqrt{1+\kappa _{0}\theta },\qquad b=\sqrt{1-\kappa _{0}p_{0}},
\end{equation}%
respectively and they must satisfy an equation of state of the form $a=a(b)$.
The mass continuity and the hydrostatic equilibrium equations must be
integrated with the boundary conditions
\begin{eqnarray}
m(0) &=&0,\qquad \theta (0)=1,  \notag \\
b(0) &=&\sqrt{1-\kappa _{0}p_{0c}},\qquad b\left( \eta _{S}\right) =1,
\end{eqnarray}%
where $p_{0c}=p_{c}/\rho _{c}c^{2}$, with $p_{c}$ the central pressure,
while $\eta _{S}$ determines the radius $R$ of the star through the
condition $p(R)=0$. Once the dimensionless parameters $\left( \eta
,m_{0},\kappa _{0},\theta \right) $ are obtained from the numerical
integration of the structure equations of the star, the physical parameters
in the $q$-metric can be obtained as
\begin{eqnarray}
&&r=3.276\times 10^{6}\times \left( \frac{\rho }{\rho _{n}}\right)
^{-1/2}\times \eta \;\mathrm{cm},  \notag \\
&&m=22.107\times \left( \frac{\rho }{\rho _{n}}\right) ^{-1/2}\times
m_{0}\times M_{\odot },  \notag \\
&&\kappa =2.684\times 10^{12}\times \kappa _{0}\;\mathrm{cm^{2}},
\end{eqnarray}%
where $\rho _{n}=2\times 10^{14}$ g/cm$^{3}$ is the nuclear density, and $%
M_{\odot }=2\times 10^{33}$ g is the solar mass.

In the $g$-metric the physical mass $M(r)$ of the star is defined with the
help of the metric tensor component $e^{-\lambda }$ as
\begin{equation}
e^{-\lambda }=1-\frac{2GM(r)}{c^{2}r}.
\end{equation}%
Therefore we obtain the following relation between the masses $M(r)$ and $%
m(r)$ in the physical $g$ and auxiliary $q$ metrics,
\begin{equation}
\frac{2GM(r)}{c^{2}r}=1-\left[ 1-\frac{2Gm(r)}{c^{2}r}\right] ab.  \label{Mm}
\end{equation}

Taking into account the dimensionless variables introduced in Eqs.~(\ref{var}) we have
\begin{equation}
\frac{2M_{0}(\eta )}{\eta }=1-\left[ 1-\frac{2m_{0}(\eta )}{\eta }\right]
\sqrt{\left( 1+\kappa _{0}\theta \right) \left( 1-\kappa _{0}p_{0}\right) }.
\end{equation}

In the case of true vacuum, $a=b=1$, from Eqs. (\ref{g1})-(\ref{x1}) we
obtain the metric function $f\left( r\right) =r^{2}$, the $g$-metric
coefficients
\begin{equation}
e^{\nu \left( r\right) }=e^{-\lambda \left( r\right) }=1-\frac{2GM}{c^{2}r},
\label{a6}
\end{equation}%
and the $q$-metric coefficients,
\begin{equation}
e^{\beta \left( r\right) }=e^{-\alpha \left( r\right) }=1-\frac{2Gm}{c^{2}r},
\label{a7}
\end{equation}%
respectively, which is the Schwarzschild solution. From Eq. (\ref{Mm}), we
obtain the relation $M=m$. Therefore it follows that the physical $g$-metric is
identical to the apparent $q$ metric. Hence the EiBI theory is completely
equivalent to standard general relativity in true vacuum.

\section{High density compact objects in EiBI gravity}
\label{sect4}

In the present Section, we consider four cases of stellar structures in the
EiBI gravity model, corresponding to different choices of the equation of
state of dense matter. More specifically, we will consider the structure of
high density stars composed of matter obeying the Zeldovich (stiff fluid),
the radiation, the polytropic and the MIT bag model equations of state,
respectively. In all these cases the properties of the corresponding neutron
and quark stars are obtained by numerically integrating the structure
equations. We will compare our results with the standard general
relativistic ones, in which the structure of the high density compact
objects is described by the mass continuity and the TOV equation, given by
\begin{equation}
\frac{dM}{dr}=4\pi \rho r^{2},
\end{equation}%
and
\begin{equation}
\frac{dp}{dr}=-\frac{G\left( \rho +p/c^{2}\right) \left( M+4\pi
r^{3}p/c^{2}\right) }{r^{2}\left( 1-2GM/c^{2}r\right) },
\end{equation}%
respectively. In the dimensionless variables introduced in Eqs.~(\ref{var}),
the standard general relativistic mass continuity and TOV equations take the
form
\begin{equation}
\frac{dM_{\ast }}{d\eta }=2\theta \eta ^{2},
\end{equation}%
and
\begin{equation}
\frac{dp_{\ast }}{d\eta }=-\frac{\left( \theta +p_{\ast }\right) \left(
2p_{\ast }\eta ^{3}+M_{\ast }\right) }{\eta ^{2}\left( 1-2M_{\ast }/\eta
\right) },
\end{equation}%
respectively, where $M=c^{3}/\sqrt{2\pi G^{3}\rho _{c}}M_{\ast }$, and $%
p=\rho _{c}c^{2}p_{\ast }$.

\subsection{Compact stars in the EiBI model obeying the Zeldovich (Stiff
Fluid) EOS}

One of the most common equations of state, which has been used extensively
to study the properties of compact objects is the linear barotropic equation
of state, $p=(\gamma -1)\rho c^{2}$, with $\gamma =\mathrm{constant}\in
\lbrack 1,2]$. The Zeldovich (stiff fluid) equation of state, corresponds to
the case $\gamma =2$. This equation of state is valid for densities
significantly higher than nuclear densities, $\rho >10\rho _{n}$. It can be
obtained by constructing a relativistic Lagrangian that allows bare nucleons
to interact attractively via scalar meson exchange, and repulsively via the
exchange of a more massive vector meson \cite{60}. In the non-relativistic
limit, in both the quantum and classical theories the interaction is
mediated via Yukawa-type potentials. The vector meson exchange dominates at
the highest matter densities and, by using a mean field approximation, it
follows that in the extreme limit of infinite densities the pressure tends
to the energy density, $p\rightarrow \rho c^{2}$ \cite{60}. In this case,
the speed of sound approaches the velocity of light, i.e., $%
c_{s}^{2}=dp/d\rho \rightarrow c^{2}$, and therefore the stiff fluid
equation of state satisfies the causality condition, with the speed of sound
equal to the speed of light.

In the dimensionless variables given by Eqs.~(\ref{dim}), we obtain the following expressions
\begin{eqnarray}
&&a=\sqrt{1+\kappa _{0}\theta },\qquad b=\sqrt{1-\kappa _{0}\theta },\qquad
a^{2}=2-b^{2},  \notag \\
&&a^{2}-b^{2}=2\left( 1-b^{2}\right) ,\qquad c_{q}^{2}=-\frac{b}{\sqrt{%
2-b^{2}}}.
\end{eqnarray}%
In order to have a real $b$, the parameter $\kappa _{0}$ must satisfy the
constraint $\kappa _{0}<1$.

\begin{figure*}[!ht]
\centering \includegraphics[scale=0.70]{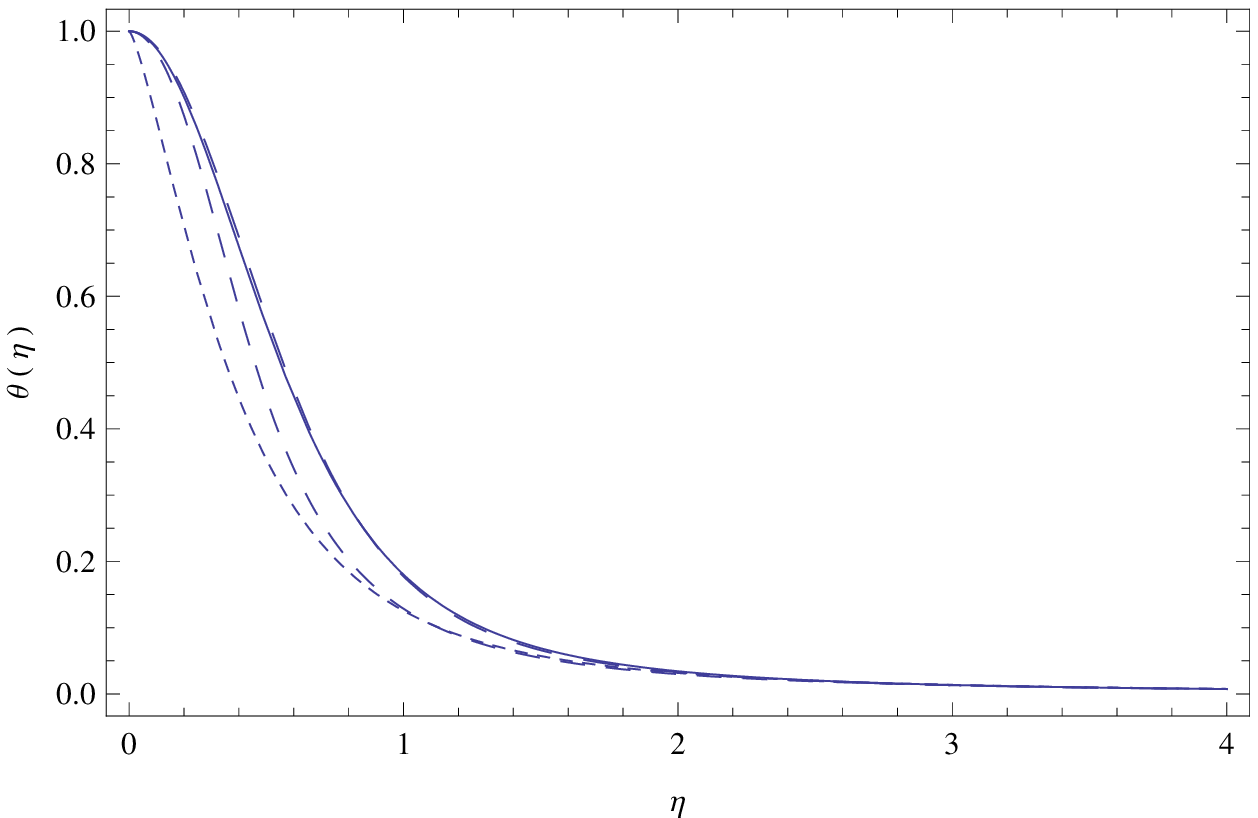}\hfill %
\includegraphics[scale=0.70]{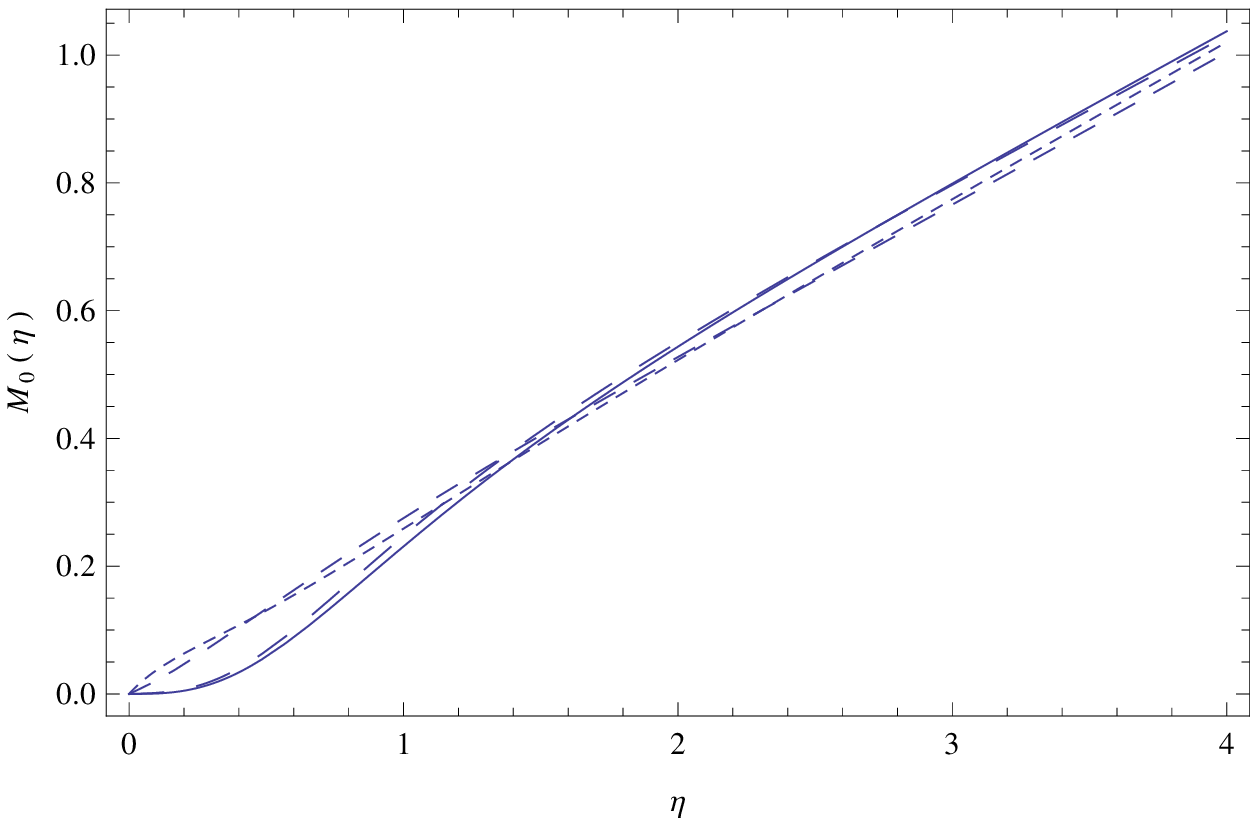}
\caption{Comparison of the dimensionless density (left figure) and $g$%
-metric mass (right figure) profiles for standard general relativistic and
EiBI gravity models with stiff fluid equation of state, for different values
of the parameter $\protect\kappa _0$: $\protect\kappa _0\rightarrow 0$ --
the general relativistic case -- (solid curve), $\protect\kappa _0=0.9999$
(dotted curve), $\protect\kappa _0=0.8$ (dashed curve), and $\protect\kappa %
_0=0.2$ (long dashed curve), respectively. The initial conditions used for
the numerical integration of the mass continuity and hydrostatic equilibrium
equations are $\protect\theta (0)=1$, $M_*(0)=0$, $m_0(0)=0$, and $b(0)=%
\protect\sqrt{1-\protect\kappa _0}$, respectively. See the text for more details.}
\label{fig1}
\end{figure*}

Then the mass continuity and the hydrostatic equilibrium equation for the
stiff fluid star in EiBI gravity become
\begin{equation}
\frac{dm_{0}}{d\eta }=\frac{2}{\kappa _{0}}\left( 1+\frac{1-2b^{2}}{b^{3}%
\sqrt{2-b^{2}}}\right) \eta ^{2},
\end{equation}
and
\begin{eqnarray}
\frac{db}{d\eta } &=&\frac{\left( b^{2}-1\right) }{\kappa _{0}b^{2}}\times
\notag \\
&& \hspace{-1.25cm} \times \frac{\left[ 2\left( b^{5}-2b^{3}+\sqrt{2-b^{2}}%
\right) \eta ^{3}-b^{3}\left( b^{2}-2\right) \kappa _{0}m_{0}\right] }{%
\left( 2b^{2}-3\right) \eta \left( \eta -2m_{0}\right) },
\end{eqnarray}
respectively. The variation of the density and mass profiles of the stiff
fluid stars in standard general relativity and EiBI gravity model are
represented in Fig.~\ref{fig1}.

As one can see from the figures, there is a very good concordance between
the general relativistic and the EiBI gravity model predictions. For values
of $\kappa $ so that $\kappa \leq 0.2$ basically the predictions of the two
models coincide. For values of $\kappa $ in the range of $\kappa \in
(0.3,0.9999)$ there are some small quantitative differences in the density
and mass profiles, but which do not lead to significant differences in the
global astrophysical parameters (mass and radius) of the star. In standard
general relativity the maximum mass of neutron stars was obtained in \cite%
{Ruff}, and estimated to be of the order of $3.2M_{\odot }$, by assuming
that at densities higher than $4.6\times 10^{14}$ g/cm$^{3}$ the equation of
state of matter is the stiff fluid equation of state. The dimensionless
density $\theta \approx 0$ at $\eta _{S}\approx 2$, corresponds to a
dimensionless mass value of $M_{0}\approx 0.5$. Hence we obtain the radius
and the mass of the star as a function of the central density in the form
\begin{equation}
R\approx \frac{9.268\times 10^{13}}{\sqrt{\rho _{c}}},\qquad M(R)\approx
\frac{1.563\times 10^{8}}{\sqrt{\rho _{c}}}.
\end{equation}%
For central densities of the order of $\rho _{c}=2\times 10^{15}$g/cm$^{3}$,
the radius and the mass of the neutron star are $R=2.07\times 10^{6}$ cm,
and $M=3.49M_{\odot }$, respectively. Hence the EiBI gravity corrections do
not modify significantly the maximum values of the static neutron star
masses.

\subsection{Compact star with a radiation equation of state in EiBI gravity}

For a radiation-type high density fluid the equation of state (EOS) is $%
p=\rho c^{2}/3$ \cite{60}. For this case we get
\begin{equation}
a=\sqrt{1+\kappa _{0}\theta },\qquad b=\sqrt{1-\frac{\kappa _{0}\theta }{3}},
\end{equation}%
with the parameter $\kappa _{0}$ satisfying the constraint $\kappa _{0}<3$,
thus we obtain the following results
\begin{eqnarray}
a^{2}=4-3b^{2}, &&\quad a^{2}-b^{2}=4\left( 1-b^{2}\right) ,  \notag \\
c_{q}^{2} &=&-\frac{3b}{\sqrt{4-3b^{2}}}.
\end{eqnarray}

\begin{figure*}[th]
\centering \includegraphics[scale=0.70]{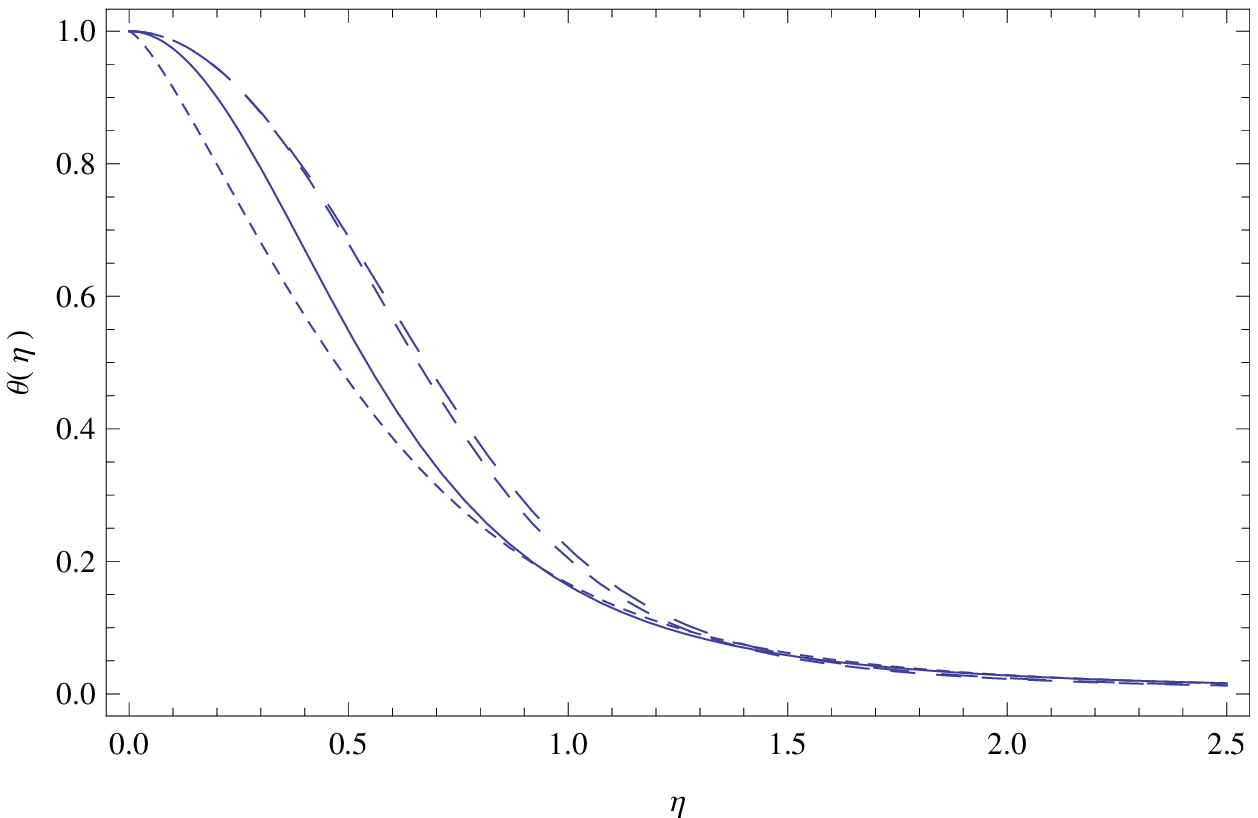} %
\includegraphics[scale=0.70]{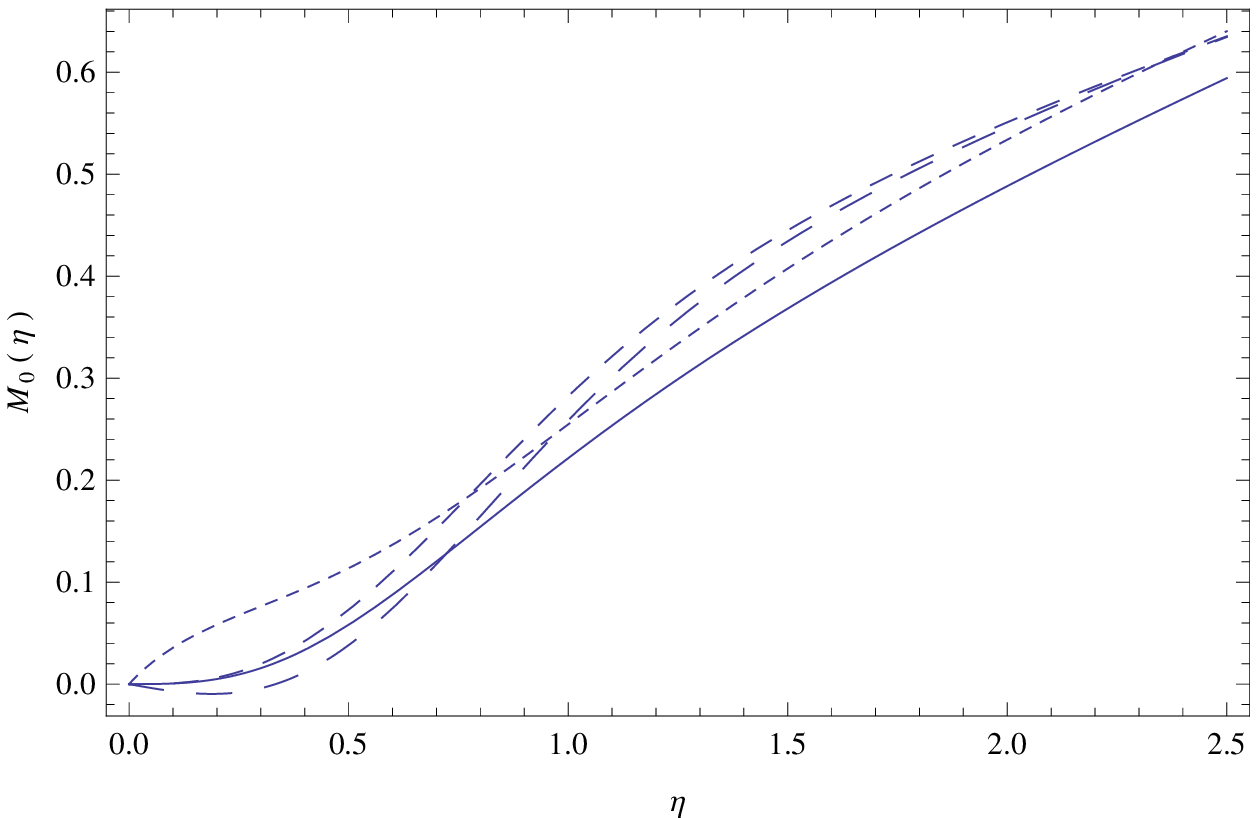}
\caption{The plots depict the comparison of the dimensionless density (left
figure) and $g$-metric mass (right figure) profiles for standard general
relativistic and EiBI gravity models with radiation fluid equation of state,
for different values of the parameter $\protect\kappa _{0}$: $\protect\kappa %
_{0}\rightarrow 0$ -- the general relativistic case -- (solid curve), $%
\protect\kappa _{0}=2.9999$ (dotted curve), $\protect\kappa _{0}=2$ (dashed
curve), and $\protect\kappa _{0}=1$ (long dashed curve), respectively. The
initial conditions used for the numerical integration of the mass continuity
and hydrostatic equilibrium equations are $\protect\theta (0)=1$, $M_{\ast
}(0)=0$, $m_{0}(0)=0$, and $b(0)=\protect\sqrt{1-\protect\kappa _{0}/3}$,
respectively. We refer the reader to the text for further details.}
\label{fig2}
\end{figure*}

For a high density star with a radiation-like EOS, the mass continuity and
the hydrostatic equilibrium equations take the form
\begin{equation}
\frac{dm_{0}}{d\eta }=\frac{2}{\kappa _{0}}\left( 1+\frac{2-3b^{2}}{b^{3}%
\sqrt{4-3b^{2}}}\right) \eta ^{2},
\end{equation}%
and
\begin{eqnarray}
\frac{db}{d\eta } &=&\frac{2\sqrt{4-3b^{2}}\left( b^{2}-1\right) }{\kappa
_{0}b^{2}}\times  \notag \\
&&\hspace{-1.35cm}\times \frac{\left[ 2\left( b^{3}\sqrt{4-3b^{2}}%
+b^{2}-2\right) \eta ^{3}-b^{3}\sqrt{4-3b^{2}}\kappa _{0}m_{0}\right] }{%
\left( 3b^{4}-14b^{2}+12\right) \eta \left( \eta -2m_{0}\right) },
\end{eqnarray}%
respectively. The variation of the dimensionless density and mass profiles
for the radiation-type equation of state is represented in Fig.~\ref{fig2}.

For a radiation fluid like star, the qualitative behavior of the mass and of
the density are similar in both general relativity, and EiBI gravity.
However, some quantitative differences between the two models do appear for
this case. The dimensionless density reaches the value zero at around $\eta
_{S}=2$, $\theta \left( \eta _{S}\right) \approx 0$, with the corresponding
dimensionless masses being given by $M_{\ast }\approx 0.45$ in general
relativity, and by $M_{0}\approx 0.55$ in EiBI gravity, corresponding to the
radii and masses
\begin{eqnarray}
R_{GR}\approx R_{EiBI} &\approx &\frac{9.268\times 10^{13}}{\sqrt{\rho _{c}}},
  \notag \\
M_{GR}(R)\approx \frac{1.40\times 10^{8}}{\sqrt{\rho _{c}}}, &&\quad
M_{EiBI}(R)\approx \frac{1.719\times 10^{8}}{\sqrt{\rho _{c}}}.  \notag
\end{eqnarray}%
For $\rho _{c}=2\times 10^{15}$ g/cm$^{3}$, we obtain $R_{GR}\approx
R_{EiBI}\approx 2.07\times 10^{6}$ cm, $M_{GR}(R)\approx 3.24M_{\odot }$,
and $M_{EiBI}(R)\approx 3.84M_{\odot }$, representing an increase of around
22\% of the high density neutron star mass due to the EiBI gravitational
effects.

\subsection{Polytropic stars in EiBI gravity model}

The polytropic equation of state
\begin{equation}
p=K\rho ^{\Gamma }=K\rho ^{1+1/n},
\end{equation}%
where $K$, $\ \Gamma $ and $n$ are usually called the polytropic constant,
the polytropic exponent and the polytropic index respectively, has been extensively
used in astrophysics for the study of white dwarfs and neutron stars \cite%
{pol}. By introducing the transformation
\begin{equation}
\rho =\rho _{c}\theta ^{n},
\end{equation}%
the polytropic EOS can be written as
\begin{equation}
p=K\rho _{c}^{1+1/n}\theta ^{n+1}.
\end{equation}%
Therefore we obtain for the parameters $a$ and $b$ the expressions
\begin{equation}
a=\sqrt{1+\kappa _{0}\theta ^{n}},\qquad b=\sqrt{1-\kappa _{0}k_{0}\theta
^{n+1}},
\end{equation}%
where $k_{0}=K\rho _{c}^{1/n}/c^{2}$. The parameter $\kappa _{0}$ must
satisfy the constraint $\kappa _{0}<1/k_{0}$. Therefore we obtain
\begin{equation}
a=\sqrt{1+\kappa _{0}^{1/(n+1)}\left( \frac{1-b^{2}}{k_{0}}\right) ^{n/(n+1)}%
}.
\end{equation}

\begin{figure*}[th]
\centering \includegraphics[scale=0.70]{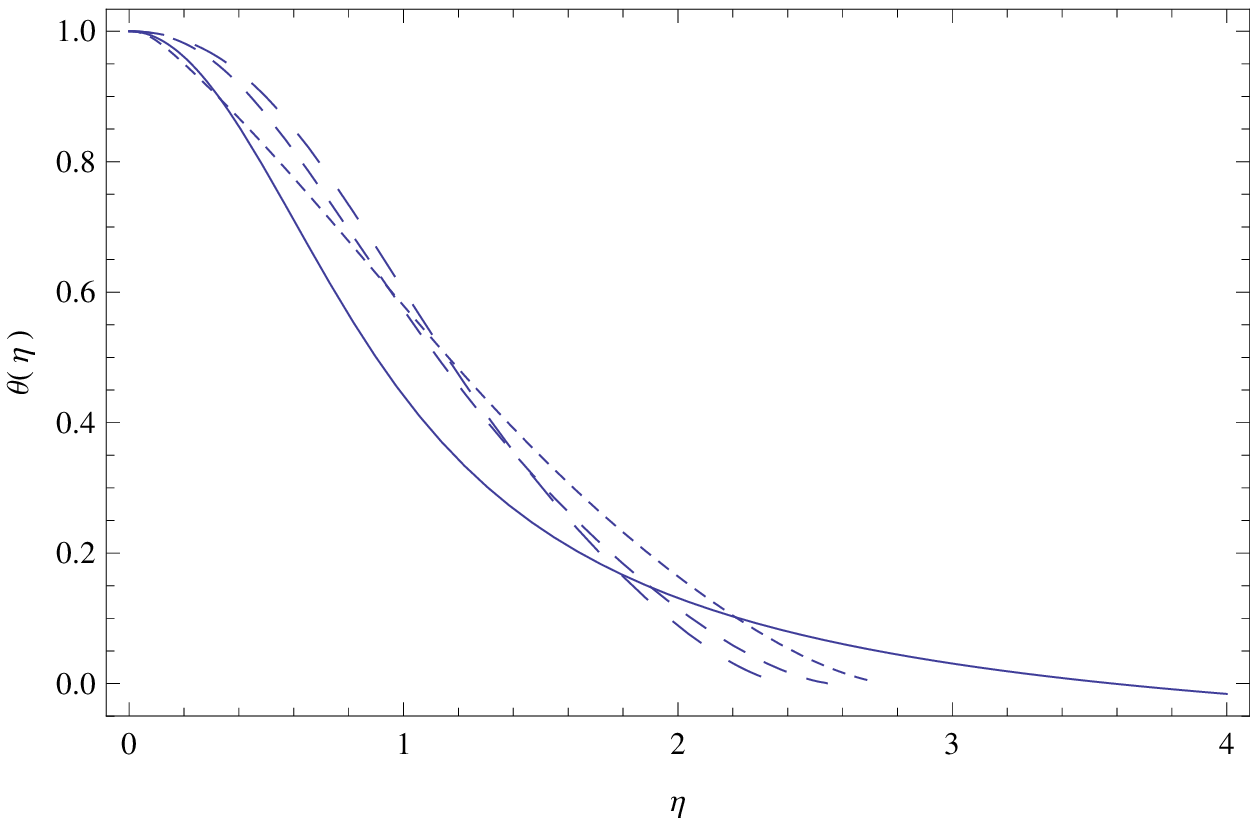} %
\includegraphics[scale=0.70]{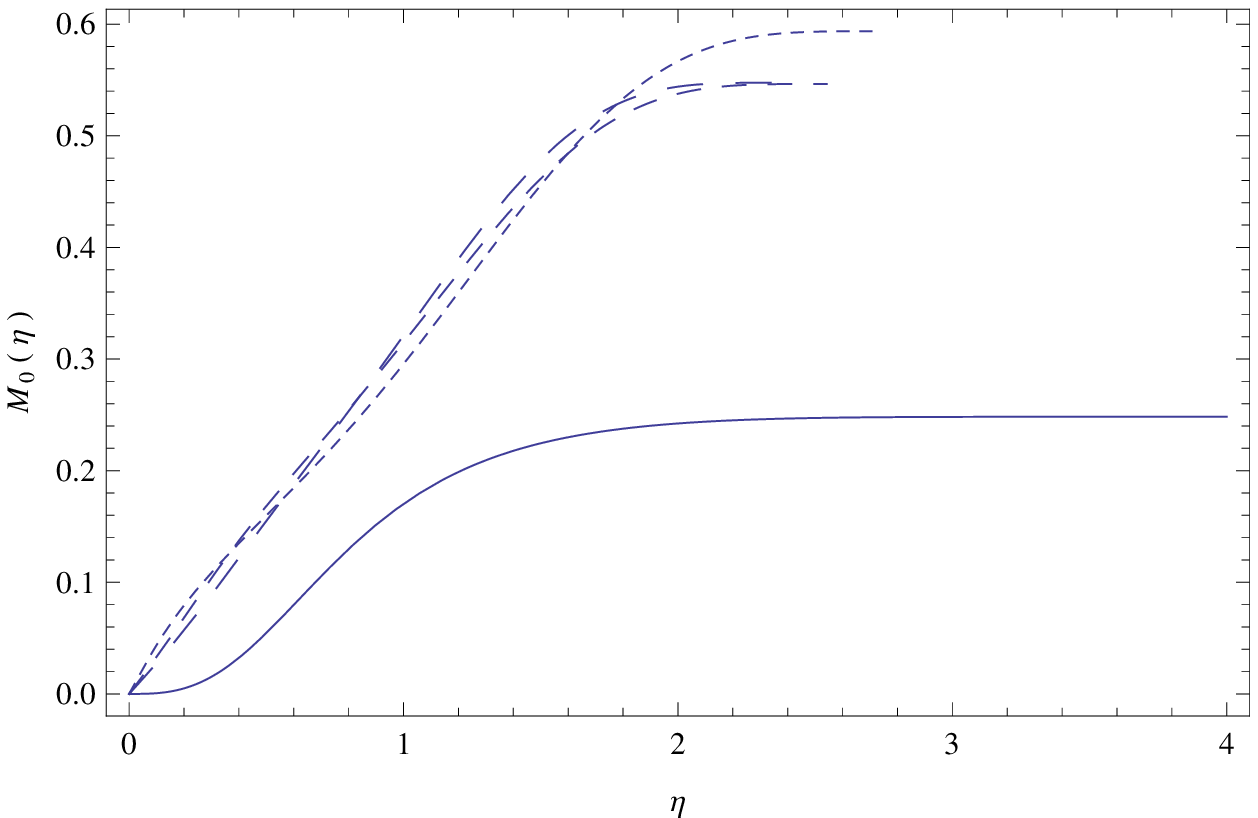}
\caption{Comparison of the dimensionless density (left figure) and $g$%
-metric mass (right figure) profiles for standard general relativistic and
EiBI gravity models with $n=3$ polytropic equation of state, for $k_{0}=0.13$%
, and for different values of the parameter $\protect\kappa _{0}$: $\protect%
\kappa _{0}\rightarrow 0$ -- the general relativistic case -- (solid curve),
$\protect\kappa _{0}=7.62$ (dotted curve), $\protect\kappa _{0}=7$ (dashed
curve), and $\protect\kappa _{0}=6.5$ (long dashed curve), respectively. The
initial conditions used for the numerical integration of the mass continuity
and hydrostatic equilibrium equations are $\protect\theta (0)=1$, $M_{\ast
}(0)=0$, $m_{0}(0)=0$, and $b(0)=\protect\sqrt{1-k_{0}\protect\kappa _{0}}$,
respectively. }
\label{fig3}
\end{figure*}

Hence the mass continuity and the hydrostatic equilibrium equations for
polytropic stars in EiBI gravity take the form
\begin{equation}
\frac{dm_{0}}{d\eta }=\frac{1}{\kappa _{0}}\left[ 2+\frac{1-3b^{2}+\kappa
_{0}^{\frac{1}{n+1}}\left( \frac{1-b^{2}}{k_{0}}\right) ^{\frac{n}{n+1}}}{%
b^{3}\sqrt{1+\kappa _{0}^{\frac{1}{n+1}}\left( \frac{1-b^{2}}{k_{0}}\right)
^{\frac{n}{n+1}}}}\right] \eta ^{2},
\end{equation}%
and
\begin{widetext}
 \begin{equation}
\frac{db}{d\eta }=\frac{2\left\{ \left[ \frac{b^{2}+1+\kappa _{0}^{\frac{1}{%
n+1}}\left( \frac{1-b^{2}}{k_{0}}\right) ^{\frac{n}{n+1}}}{b^{3}\sqrt{%
1+\kappa _{0}^{\frac{1}{n+1}}\left( \frac{1-b^{2}}{k_{0}}\right) ^{\frac{n}{%
n+1}}}}-2\right] \frac{\eta ^{3}}{\kappa _{0}}+m_{0}\right\} }{\eta
^{2}\left( 1-  \frac{2m_{0}}{\eta} \right) \left\{ \frac{bn\kappa _{0}^{\frac{1}{n+1}%
}\left( \frac{1-b^{2}}{k_{0}}\right) ^{-\frac{1}{n+1}}}{k_{0}\left(
n+1\right) \left[ 1+\kappa _{0}^{\frac{1}{n+1}}\left( \frac{1-b^{2}}{k_{0}}%
\right) ^{\frac{n}{n+1}}\right] }+\frac{4b}{1-b^{2}+\kappa _{0}^{\frac{1}{n+1%
}}\left( \frac{1-b^{2}}{k_{0}}\right) ^{\frac{n}{n+1}}}+\frac{3}{b}\right\} },
\end{equation}
 \end{widetext}respectively. For the standard general relativistic case, the
mass continuity and the hydrostatic equilibrium equations are given by
\begin{equation}
\frac{dM_{\ast }}{d\eta }=2\theta ^{n}\eta ^{2},
\end{equation}%
and
\begin{equation}
\frac{d\theta }{d\eta }=-\frac{\left( 1+k_{0}\theta \right) \left( M_{\ast
}+2k_{0}\eta ^{3}\theta ^{n+1}\right) }{k_{0}(n+1)\eta ^{2}\left( 1-2M_{\ast
}/\eta \right) },
\end{equation}%
respectively. In the following, we restrict our analysis to the case of the
degenerate ultra-relativistic neutron gas, with equation of state $p=\left(
3\pi ^{2}\right) ^{1/3}\left( \hbar c/4\right) \left( \rho /m_{n}\right)
^{4/3}$, where $m_{n}$ is the neutron mass. This equation of state has the
polytropic index $n=3$, and $K=1.23\times 10^{15}$. For central star
densities of the order of four times the nuclear density, $\rho _{c}=8\times
10^{14}$ g/cm$^{3}$, the parameter $k_{0}$ has the value $k_{0}=0.13$, which
is the value we will use for the numerical study of the polytropic stars.
For this value of $k_{0}$ we obtain for $\kappa _{0}$ the constraint $\kappa
_{0}<7.692$. The variation of the dimensionless density and mass of the
general relativistic and EiBI stars with $n=3$ polytropic equation of state
are represented in Fig.~\ref{fig3}.

In the case of the polytropic equation of state, and for the chosen values
of the physical parameters, significant differences between the global
properties of stars in the two gravitational theories appear. The
dimensionless radius $\eta _{S}$ of the star in the EiBI model varies in the
range $\eta _{S}\in (2.2,2.9)$ for $\kappa _{0}\in (6.5,7.2)$, while the
dimensionless radius of the polytropic general relativistic star is around $%
\eta _{S}\approx 4$. This shows that polytropic stars are more compact
(smaller radius) in EiBI gravity. The most importance differences arise in
the mass of the stars. While the dimensionless general relativistic mass $%
M_{0}$ is around 0.22, the dimensionless mass $M_{0}$ of the polytropic
stars in EiBI gravity is in the range of $M_{0}\approx 0.55-0.60$, which
entails that the EiBI polytropic stars have masses around 2.5 times larger
than the general relativistic ones. For the considered central density of $%
\rho _{c}=8\times 10^{14}$g/cm$^{3}$, the radius and the mass of the general
relativistic $n=3$ polytrope is $R=6.55\times 10^{6}$ cm, and $%
M_{GR}(R)=2.43M_{\odot }$, respectively, while the corresponding masses in
EiBI gravity are $E_{EiBI}\approx 4.58\times 10^{6}$ cm, and $%
M_{EiBI}\approx 6.07M_{\odot }$. Therefore EiBI gravity allows the existence
of more massive stars than standard general relativity.

\subsection{Structure and properties of quark stars in EiBI gravity}

The chemical composition of neutron stars at densities beyond the nuclear
saturation remains uncertain, with alternatives ranging from purely
nucleonic composition through hyperon or meson condensates, to deconfined
quark matter \cite{Reddy}. It was suggested that at all pressures strange
quark matter (consisting of up $u$, down $d$, and strange $s$ quarks) might
be the absolute ground state of hadronic matter \cite{Witten, Glend}.

Quark matter is formed from a Fermi gas of $3A$ quarks, constituting a
single color singlet baryon with baryon number $A$. The theory of the
equation of state of strange matter is directly based on the fundamental
Quantum Chromodynamics (QCD) Lagrangian \cite{Wein}. In first order
perturbation theory, by neglecting the quark masses, the equation of state for
zero temperature quark matter is given by the MIT Bag model equation of
state \cite{Witten, Glend, Wein}
\begin{equation}
p=\frac{1}{3}\left( \rho -4B\right) c^{2},  \label{bag}
\end{equation}%
where $B$ is the difference between the energy density of the perturbative
and non-perturbative QCD vacuum (the bag constant). Equation~(\ref{bag}) is
essentially the equation of state of a gas of massless particles with
corrections due to the QCD trace anomaly and perturbative interactions. The
vacuum pressure $B$, which holds quark matter together, is a simple model
for the long-range, confining interactions in QCD. At the surface of the
quark star, as $p\rightarrow 0$, we have $\rho \rightarrow 4B$. The typical
value of the bag constant is of the order $B\approx 10^{14}$ g/cm$^{3}$ \cite%
{Witten}. After the neutron matter-quark matter phase transition (which is
supposed to take place in the dense core of neutron stars) the energy
density of strange matter is $\rho \approx 5\times 10^{14}$ g/cm$^{3}$.
Therefore quark matter always satisfies the condition $p\geq 0$. In the
dimensionless variables introduced by Eqs.~(\ref{dim}) the Bag model
equation of state takes the form
\begin{equation}
p_{0}=\frac{1}{3}\left( \theta -4B_{0}\right) ,
\end{equation}%
where $B_{0}=B/\rho _{c}$. For the parameters $a$ and $b$ we obtain
\begin{equation}
a=\sqrt{1+\kappa _{0}\theta },\qquad b=\sqrt{1-\frac{\kappa _{0}}{3}\left(
\theta -4B_{0}\right) },
\end{equation}%
which provides the following relationships
\begin{eqnarray}
a &=&\sqrt{4\left( 1+\kappa _{0}B_{0}\right) -3b^{2}}, \\
c_{q}^{2} &=&-\frac{3b}{\sqrt{4\left( 1+\kappa _{0}B_{0}\right) -3b^{2}}}, \\
a^{2}-b^{2} &=&4\left( 1+\kappa _{0}B_{0}-b^{2}\right) ,
\end{eqnarray}%
respectively. The parameter $\kappa _{0}$ must satisfy the constraint $%
\kappa _{0}<3/\left( 1-4B_{0}\right) $. Therefore, the gravitational field
equations describing the structure of a quark star satisfying the MIT Bag
model equation of state in EiBI gravity take the form
\begin{equation}
\frac{dm_{0}}{d\eta }=\frac{2}{\kappa _{0}}\left[ 1+\frac{2\left( \kappa
_{0}B_{0}+1\right) -3b^{2}}{b^{3}\sqrt{4\left( \kappa _{0}B_{0}+1\right)
-3b^{2}}}\right] \eta ^{2},
\end{equation}%
and
\begin{widetext}
\be
\frac{db}{d\eta}=-\frac{2\left(b^2-\kappa _0B_0 -1\right) \sqrt{-3 b^2+4 \kappa _0B_0 +4} \left[-2
   \left(b^2-2\right) \eta ^3+b^3 \sqrt{-3 b^2+4\kappa _0 B_0 +4} \left(\kappa _0
   m_0-2 \eta ^3\right)+4 \kappa _0B_0 \eta ^3 \right]}{\kappa _0b^2 \eta ^2
   \left(1 -2 m_0/\eta \right) \left[3 b^4-14 \left(1+\kappa _0B_0 \right)b^2+12\left(1+\kappa _0B_0\right)^2\right]}.
   \ee
   \end{widetext}

For the central density of the quark star we adopt the value $\rho
_c=4\times 10^{15}$ g/cm$^3$, leading to $4B_0=0.1$, and $B_0=0.025$,
respectively. With these values for $\kappa _0$ we have the constraint $%
\kappa _0<3.33$. The variations of the density and mass profiles of the
quark stars in general relativity and EiBI gravity are presented in Fig.~\ref%
{fig4}.
\begin{figure*}[!ht]
\centering \includegraphics[scale=0.70]{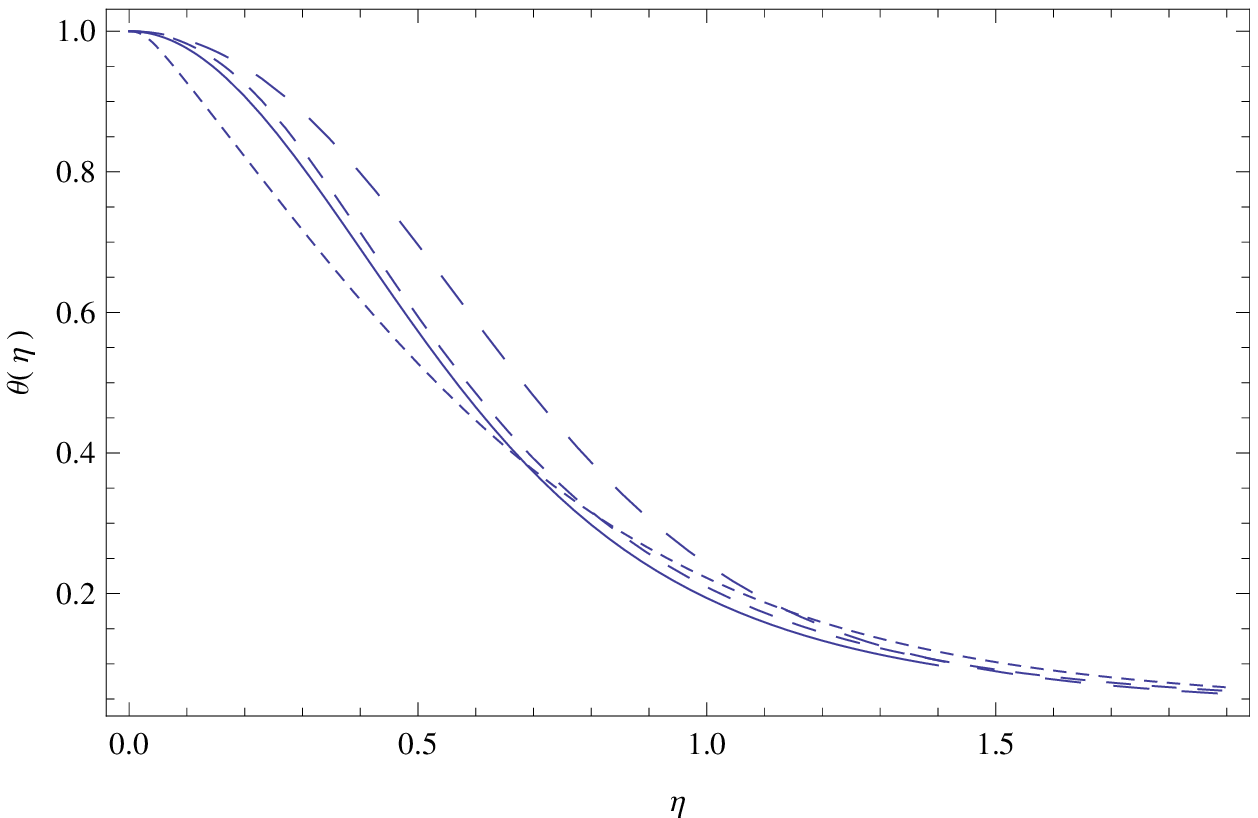} %
\includegraphics[scale=0.70]{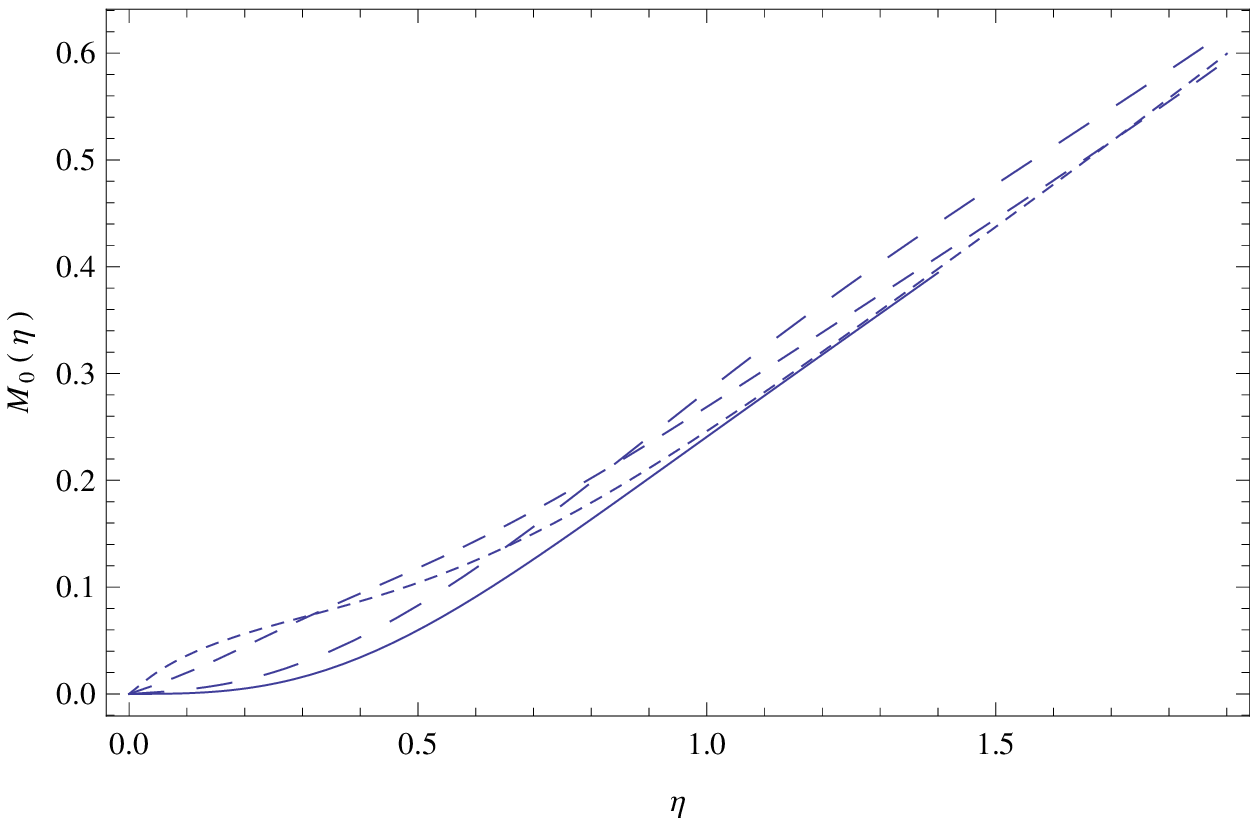}
\caption{ Dimensionless density (left figure) and $g$-metric mass (right
figure) profiles for standard general relativistic and EiBI gravity models
for quark stars with the MIT bag model equation of state, for $B_0=0.025$,
and for different values of the parameter $\protect\kappa _0$: $\protect%
\kappa _0\rightarrow 0$ -- the general relativistic case -- (solid curve), $%
\protect\kappa _0=3.32$ (dotted curve), $\protect\kappa _0=3$ (dashed curve),
and $\protect\kappa _0=2.5$ (long dashed curve), respectively. The initial
conditions used for the numerical integration of the mass continuity and
hydrostatic equilibrium equations are $\protect\theta (0)=1$, $M_*(0)=0$, $%
m_0(0)=0$, and $b(0)=\protect\sqrt{1-\protect\kappa _0\left(1-4B_0\right)/3}$,
respectively. See the text for details.}
\label{fig4}
\end{figure*}

The quark star models are relatively similar in both general relativity and
EiBI gravity. The dimensionless radius of the quark star is $\eta _S\approx
1.3$, obtained from the condition $\theta \left(\eta _S\right)=0.10$, with
the corresponding quark dimensionless mass of $M_*\approx 0.3$ and $%
M_0\approx 0.38$, corresponding to $\kappa _0=3.2$. The general relativistic
quark star radius is $R=9.52\times 10^5$ cm, while its mass is around $%
M_{GR}(R)=1.48M_{\odot}$. The EiBI star has a similar radius, and a mass
given by $M_{EiBI}\approx 1.87M_{\odot}$. Similarly to the polytropic case,
EiBI gravity effects lead to an increase of around 25\% of the mass of the
quark stars described by the MIT bag model equation of state.

\section{An exact stellar solution in EiBI gravity: the $a^{2}=3b^{2}$ case}
\label{sect5}

Due to the highly nonlinear nature of the EiBI gravitational field equations, it is extremely difficult to obtain exact analytical solutions. However, in this section we present an exact solution for a peculiar compact object.
In the following, we adopt for simplicity the natural system of units with $%
G=c=1$. We consider the specific case
\begin{equation}
a^{2}=3b^{2}.
\end{equation}%
Then, the corresponding relation between the energy density and the pressure
is
\begin{equation}
p=-\frac{1}{3}\rho +\frac{1}{12\pi \kappa }.  \label{EoS-string}
\end{equation}

In this case, the field equations are essentially simplified, and admit an
exact solution. In particular, Eq.~(\ref{g1}) can be integrated to give
\begin{equation}
e^{-\alpha }=1-\frac{2m(r)}{r},
\end{equation}%
where
\begin{equation}
m(r)=M+\frac{r^{3}}{6\kappa },
\end{equation}%
and $M$ is an arbitrary constant of integration. The regularity condition
requires that $m(0)=0$, hence we assume that $M=0$. Therefore we obtain the
metric function
\begin{equation}
e^{-\alpha }=1-\frac{r^{2}}{3\kappa }.  \label{alpha-string}
\end{equation}

The conservation relation (\ref{L}) now yields the result
\begin{equation}
\frac{d\beta }{dr}=\frac{4}{b}\frac{db}{dr}.  \label{beta-string}
\end{equation}%
Using the relation $a^{2}=3b^{2}$ and substituting Eqs.~(\ref{alpha-string})
and (\ref{beta-string}) into Eq.~(\ref{g2}) yields the following equation
for $b(r)$,
\begin{equation}
3\kappa \left( 1-\frac{r^{2}}{3\kappa }\right) \frac{db^{2}}{dr}+rb^{2}=%
\sqrt{3}\,r,
\end{equation}%
with the general solution given by
\begin{equation}
b^{2}(r)=\sqrt{3}-C\sqrt{1-\frac{r^{2}}{3\kappa }},
\end{equation}%
where $C$ is an arbitrary constant of integration. In order to fix $C$, we
assume that $p(0)=p_{c}$, where $p_{c}$ is the pressure at the center of the
star. Then, using the relation $b^{2}=1-8\pi \kappa p$, we verify that the
arbitrary constant of integration is given by
\begin{equation*}
C=8\pi \kappa p_{c}+\sqrt{3}-1,
\end{equation*}%
and hence
\begin{equation}
b^{2}(r)=\sqrt{3}-\left( 8\pi \kappa p_{c}+\sqrt{3}-1\right) \sqrt{1-\frac{%
r^{2}}{3\kappa }}\,.
\end{equation}

The corresponding relation for $p(r)$ takes the form
\begin{equation}
p(r)=\frac{1}{8\pi \kappa }\left[ \left( 8\pi \kappa p_{c}+\sqrt{3}-1\right)
\sqrt{1-\frac{r^{2}}{3\kappa }}-\sqrt{3}+1\right] .  \label{p-string}
\end{equation}%
The radius of the star is defined as a sphere $r=R$ where the pressure is
equal to zero, i.e. $p(R)=0$. From Eq. (\ref{p-string}) we find
\begin{equation}
R^{2}=\frac{48\pi \kappa ^{2}p_{c}(4\pi \kappa p_{c}+\sqrt{3}-1)}{(8\pi
\kappa p_{c}+\sqrt{3}-1)^{2}}.
\end{equation}

Note that $R$ depends on two parameters $\kappa $ and $p_{c}$, i.e. $%
R=R(\kappa ,p_{c})$. It will be useful to consider various limiting cases.
Namely, by fixing the value of $p_{c}$, then $R\rightarrow 0$ if $\kappa
\rightarrow 0$, and $R\rightarrow \infty $ if $\kappa \rightarrow \infty $.
Now, by fixing the value of $\kappa $, then $R\rightarrow 0$ if $%
p_{c}\rightarrow 0$, and $R\rightarrow R_{\mathrm{max}}=\sqrt{3\kappa }$ if $%
p_{c}\rightarrow \infty $. It is worth noticing that the size of a star
supported by this exact solution model cannot exceed the maximal size $R_{%
\mathrm{max}}=\sqrt{3\kappa }$. Finally, note that $p(r)$ given by Eq. (\ref%
{p-string}) is monotonically decreasing from $p_{c}$ to $0$ within the
interval $r\in \lbrack 0,R]$.

Using Eq.~(\ref{beta-string}) yields the metric function
\begin{equation}
e^{\beta (r)}=e^{\beta _{c}}\frac{b^{4}(r)}{b_{c}^{4}},
\end{equation}%
where $\beta _{c}=\beta (0)$ and $b_{c}=b(0)$. Then, the explicit
expressions for $\alpha (r)$ and $b(r)$ are found. Using the relation $%
a^{2}=3b^{2}$ and Eqs.~(\ref{x1}), we can easily obtain the following metric
functions
\begin{eqnarray}
e^{\nu (r)} &=&\frac{\sqrt{3}e^{\beta _{c}}}{b_{c}^{4}}\left[ \sqrt{3}%
-\left( 8\pi \kappa p_{c}+\sqrt{3}-1\right) \sqrt{1-\frac{r^{2}}{3\kappa }}%
\right] , \\
e^{\lambda (r)} &=&\frac{1}{\sqrt{3}}\left( 1-\frac{r^{2}}{3\kappa }\right)
^{-1}\times  \notag \\
&&\left[ \sqrt{3}-\left( 8\pi \kappa p_{c}+\sqrt{3}-1\right) \sqrt{1-\frac{%
r^{2}}{3\kappa }}\right] ^{-1}, \\
f(r) &=&\frac{r^{2}}{\sqrt{3}}\left[ \sqrt{3}-\left( 8\pi \kappa p_{c}+\sqrt{%
3}-1\right) \sqrt{1-\frac{r^{2}}{3\kappa }}\right] ^{-1}.
\end{eqnarray}

The resulting $g$-metric has the following form
\begin{equation}  \label{metric-string}
ds^{2}=-\frac{\sqrt{3}e^{\beta _{c}}b^{2}(r)}{b_{c}^{4}}dt^{2}+\frac{1}{%
\sqrt{3}b^{2}(r)}\left[ \frac{dr^{2}}{1-\frac{r^{2}}{3\kappa }}+r^{2}d\Omega
^{2}\right].
\end{equation}
We stress that $r<R_{\mathrm{max}}=\sqrt{3\kappa }$, therefore the term $%
1-r^{2}/3\kappa $ in the line element (\ref{metric-string}) is strictly
positive.

Using the equation of state (\ref{EoS-string}) yields
\begin{equation}
\rho (r)=-3p(r)+\frac{1}{4\pi \kappa }.
\end{equation}
Since $p(r)$ is monotonically decreasing from $p_{c}$ to $0$ within the
interval $r\in \lbrack 0,R]$, the energy density $\rho (r)$ is monotonically
increasing from $\rho _{c}=-3p_{c}+1/4\pi \kappa $ to $1/4\pi \kappa $.
Demanding the positivity of the energy density, i.e., $\rho >0$ yields the
following condition for $p_{c}$
\begin{equation}
p_{c}<\frac{1}{12\pi \kappa }.
\end{equation}%
Therefore, the pressure at the center of the star is restricted by the value
$p_{\mathrm{max}}=1/12\pi \kappa $.

\section{Conclusions}
\label{sect6}

In the present paper, we have considered the properties of specific stellar
models in the recently proposed EiBI gravity model, and we have performed a
comparative study of high density compact objects in standard general
relativity and EiBI gravity, respectively. Generally, on a qualitative
level, the predictions of these two theoretical models do agree, but
important quantitative differences also appear. An important and interesting
feature of EiBI gravity is that it predicts more massive objects than
general relativity, with an equation of state dependent increase in the
stellar mass in the range of 22\%-26\%. In the analysis outlined in this
work, we have also restricted our study to positive values of the parameter $%
\kappa _{0}$, since only in this case more massive stellar configurations
than the general relativistic ones are possible. By assuming a linear
equation of state of the form $\rho +3p=1/4\pi \kappa $, we have also
obtained the complete exact analytical solution of the gravitational field
equations describing the interior of an ``exotic'' star in EiBI gravity. In
the limit of large parameter $\kappa \rightarrow \infty $, the obtained
solution describes the string gas stellar model, with the equation of state
of the form $\rho +3p=0$ \cite{Alex}. Note that a short review on the
building blocks of the string gas cosmological model has been recently
presented in \cite{Robert}.

One of the fundamental property of stellar models is their stability with
respect to small perturbations. The stability of EiBI compact stars was
considered in \cite{Lin1}, where it was shown that the standard results of
stellar stability theory still hold in this theory. For the maximum-mass
stellar configuration the frequency square of the fundamental oscillation
mode vanishes. However, an interesting difference with respect to general
relativity is that the criterion $dM/d\rho _{c}$ does not guarantee
stability.

A very intriguing type of astrophysical objects are stellar mass black
holes, with masses in the range of $3-6M_{\odot}$. The stellar mass black
holes have been observed in close binary systems, in which transfer of
matter from a companion star to the black hole occurs. The energy released
in the fall heats up the matter to temperatures of several hundred million
degrees, and it is radiated in $X$-rays, thus allowing the detection of the
black hole, whereas the companion star can be observed with optical
telescopes. It is estimated that in the Milky Way alone there should be at
least 1000 dormant black hole $X$ Ray Transients, while the total number of
stellar mass black holes (isolated and in binaries) could be as large as 100
million \cite{Romani}. Since ordinary neutron or quark stars in EiBI gravity
can acquire larger masses than the general relativistic maximum mass of $%
3.2M_{\odot}$, stellar mass black holes, with masses in the range of $%
3.8M_{\odot}$ and $6M_{\odot}$, respectively, could in fact be EiBI neutron
or quark stars. EiBI stars can achieve much higher masses than standard
neutron stars, thus making them possible stellar mass black hole candidates.



\section*{Acknowledgments}

\textit{Acknowledgments.} FSNL acknowledges financial support of the Funda%
\c{c}\~{a}o para a Ci\^{e}ncia e Tecnologia through the grants
CERN/FP/123615/2011 and CERN/FP/123618/2011. SVS acknowledges financial
support of the Russian Foundation for Basic Research through grants No.
11-02-01162 and 13-02-12093.

\end{document}